# Defect screening and load transfer in minimal hard–soft double networks

Fucheng Tian[1*], Feixue Lu[2], Katsuhiko Sato[3], Liangbin Li[4], Bin Li[5], Jian Ping Gong [1,2,6†]

[1] Faculty of Advanced Life Science, Hokkaido University, Sapporo 001-0021, Japan.

[2] Institute for Chemical Reaction Design and Discovery (WPI-ICReDD), Hokkaido University, Sapporo 001-0021, Japan.

[3] Program of Mathematics and Informatics, Science, University of Toyama, Toyama 930-8555, Japan.

[4] National Synchrotron Radiation Laboratory, State Key Laboratory of Advanced Glass Materials, Anhui Provincial Engineering Research Center for Advanced Functional Polymer Films, University of Science and Technology of China, Hefei, Anhui 230029, China.

[5] Department of Mechanical Engineering, Guangdong Technion-Israel Institute of Technology, Shantou, 515063, China.

[6] Co-Creation Core for Soft Materials Aspiring Research & Translation (C[3]-SMART), Hokkaido University, Sapporo 001-0021, Japan

Double network (DN) materials exhibit anomalous strength and toughness that far exceed the sum of their constituents. While widely exploited, the fundamental physical mechanisms underlying this synergy remain elusive. Here, we show that a minimal three-dimensional model of two coupled, disordered linear-elastic networks is sufficient to capture the essential physics of DN nonlinear mechanics. The model reproduces the full suite of unique mechanical behaviors, including yielding, necking, strain hardening, and the brittle-to-ductile transition. Mechanical contrast between the hard and soft networks drives inter-network load transfer, which screens defects and suppresses stress concentrations in the hard network. By defining a stress-concentration factor, $K_{sc}$, we find that the hard-network failure strain scales universally as $1/K_{sc}$, directly bridging microscopic defect screening to macroscopic yielding. We further show that complete defect screening triggers the shift from localized necking to delocalized damage. Furthermore, the stable necking plateau is identified as an energetic selection governed by the balance between potential energy release and irreversible dissipation. These findings reveal that a simple linear-elastic framework can account for the rich nonlinear landscape of DN materials, providing a general principle for designing next-generation tough solids.

* Correspondence author: tianfc@sci.hokudai.ac.jp
† Correspondence author: gong@sci.hokudai.ac.jp

A hard single polymer network containing random defects is intrinsically prone to stress concentration under loading [1], where rupture of only a few critical bonds can rapidly trigger avalanche-like brittle failure. Embedding this fragile skeleton in a suitably soft, stretchable matrix, however, yields a composite that combines strength and toughness beyond the reach of its constituents [2,3]. This hard–soft double-network (DN) strategy has emerged as a paradigmatic route for designing tough soft materials [4], exemplified by DN hydrogels and elastomers [5-8]. Its remarkable success has prompted experimental [9-15], numerical [16-22], and theoretical studies [23-27] across multiple scales that probe the mechanisms underlying toughening.

Experiments have revealed that DN toughening follows a generic sacrificial-bond mechanism [3,6,10,28,29], in which the hard network bond ruptures progressively to dissipate energy while the soft network preserves connectivity and continues to carry load. The associated mechanical signatures, including pronounced yielding, necking plateau, subsequent strain hardening, and brittle-to-ductile (BTD) transition, recur across diverse chemistries, pointing to a general topological origin [30-32]. Visualization techniques using mechanophore light emission [6,33] and other mechanochemical probes [13,14,34,35] provide microscopic insights into this toughening, showing dispersed stress concentrations and extended damage zones of hard network. Molecular and coarse-grained simulations further sharpen this picture by resolving nonaffine rearrangements and inter-network load transfer, which redistribute stress and screen stress concentration around hard network defects, thereby inducing damage delocalization [16,20,22,36].

To connect these microscopic mechanisms with the global response, researchers resort to theoretical approaches beyond the molecular scale. Continuum-scale phenomenological models [37-42], although successful in reproducing key features of DN behavior such as mechanical hysteresis and necking, lack clear microphysics. Recently, mesoscale DN models offer a promising alternative [17,43-46], capable of capturing macroscopic behavior while preserving network-level structure. Notably, a simple two-dimensional (2D) model of Walker and Fielding [17] highlighted reduced stress propagation in DNs as a key mechanism for suppressing stress concentration. Yet 2D frameworks are inadequate in mimicking the realistic three-dimensional (3D) DNs [30], especially regarding inter-network relative motion. Prevailing DN models, spanning microscopic, mesoscale, and continuum levels, typically build in strong nonlinear ingredients that obscure the fundamental origins of toughening. Once these nonlinear complexities are stripped away, the essential physics of how microscopic mechanisms in generic DN systems unfold and shape the macroscopic response remains unclear.

Here, we resolve these issues by constructing a minimal 3D linear-elastic DN model that eliminates complex constitutive nonlinearities. Despite its simplicity, this framework captures the key nonlinear mechanical signatures of DN materials through the random spatial arrangement of hard and soft networks. A fundamental defect-screening mechanism, quantified by an inverse stress-concentration factor, constrains microscopic crack opening, reduces stress concentration, and determines the macroscopic yield stress. We further rationalize the necking plateau as a dissipation-regulated phase transition and clarify how inter-network load transfer and competition govern failure-mode transitions from localized necking to delocalized damage and brittle-to-ductile fracture.

Building on our previous soft–hard composite framework [30], we construct a minimal linear hard–soft double network (LHS–DN) model that abstracts DN architecture rather than reproducing a realistic polymer network. As illustrated in Fig. 1(a) (left), a bar-shaped sample is discretized into $N_{tot}$ tetrahedral elements with a characteristic mesh size $l_e$, chosen to ensure macroscopic homogeneity [47]. Two disordered networks are then generated independently on the same underlying mesh. To construct the hard network, $N_h$ elements are randomly selected and endowed with linear-elastic properties characterized by a modulus $E_h$, Poisson's ratio $\nu$, and a critical failure energy density $W_h^*$, defining a relative density $\phi_h = N_h/N_{tot} \in [0,1]$. Independently, $N_s$ elements are selected to form a soft network with $(E_s, \nu, W_s^*)$, yielding $\phi_s = N_s/N_{tot} \in [0,1]$. Statistical independence implies that elements selected by both networks physically interpenetrate and bear load in parallel, whereas those selected by neither act as topological defects or voids. Fig.1(a) (right) display intricate network slices formed by element edges for the hard (*h*), soft (*s*), and hard-soft networks (*hs*). The network contrast is quantified by two dimensionless ratios, $\gamma_E = E_s/E_h$ and $\gamma_W = W_s^*/W_h^*$. The critical failure stress and strain of linear elastic constitute satisfy $\sigma_s^* = \sqrt{\gamma_E\gamma_W}\sigma_h^*$ and $\varepsilon_s^* = \sqrt{\gamma_W/\gamma_E}\varepsilon_h^*$.

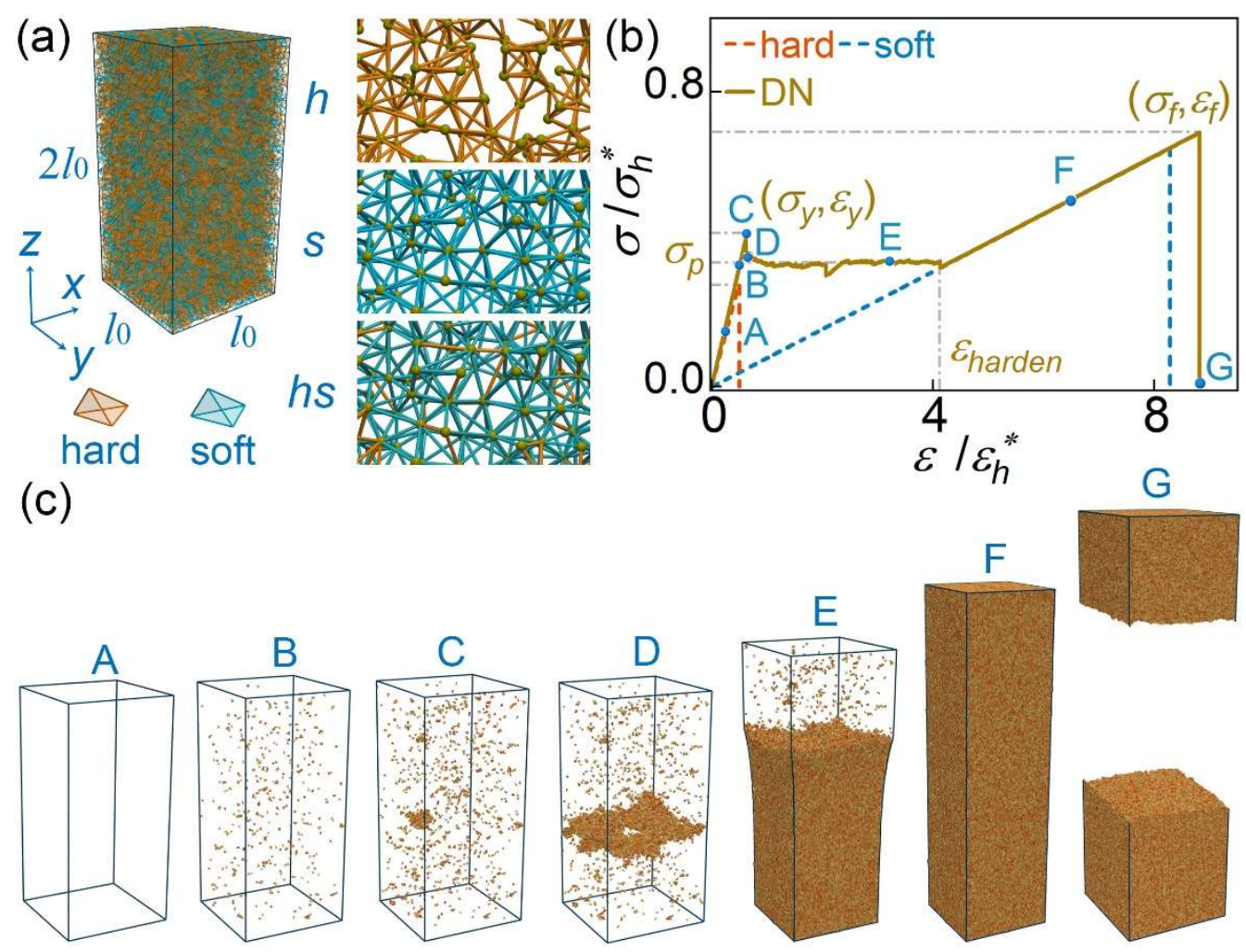

**Fig.1. Minimal 3D linear hard–soft double network (LHS-DN) model and its mechanical response. (a)** Left: schematic of the model with a sample volume $l_0 \times l_0 \times 2l_0$ discretized into tetrahedral elements, randomly assigned to hard (brown) and soft (blue) phases (see [47] for discretization and size effects). Right: representative slices highlighting the percolating hard (*h*) and soft (s) networks, and the combined (*hs*) architecture. Element edges form a disordered, interconnected framework, with nodes acting as effective crosslinks. **(b)** Stress-strain curves for the constituent single networks (dashed red: hard, dashed blue: soft) and the combined LHS-DN system (solid brown) for a specific set of $\gamma_E = 1/10$, $\gamma_W = 24$, $\phi_h = 0.6$, $\phi_s = 0.8$. **(c)** Sequential snapshots (A-G) illustrating sample deformation and the hard-network damage evolution, corresponding to the labels in (b). Brown markers indicate fractured elements.

To probe the mechanical behavior, we apply uniaxial tension along the z-axis with traction-free lateral boundaries, while pinning the symmetry axis on the top and bottom surfaces to remove rigid-body motion. Element failure is implemented via irreversible damage variables $d_\alpha \in [0,1]$. In each phase $\alpha \in [h, s]$, an element is irreversibly deactivated (i.e., $d_\alpha$ switches from 0 to 1) once its local strain-energy density exceeds the predefined threshold $W_\alpha^*$. Cascading, avalanche-like breaking triggered by stress redistribution is captured by iterating mechanical equilibrium at each fixed displacement increment until no further elements satisfy the failure criterion [47]. Unless otherwise specified, we adopt baseline parameters consistent with [30]: $E_h = 10$, $W_h^* = 0.02$, and Poison ratio $\nu = 0.35$.

A representative macroscopic response is shown in Fig. 1(b). Single-network (SN) systems fail in a defect-sensitive brittle manner, with failure stress ($\sigma_h^f$, $\sigma_s^f$) below those of their constituent elements ($\sigma_h^*, \sigma_s^*$). By contrast, the combined DN system exhibits a significantly tougher response and captures the hallmark features of DN gels [3]: an elevated yield point ($\sigma_y, \varepsilon_y$), a stable necking plateau ($\sigma_p$), and subsequent strain hardening ($\varepsilon_{harde}$ ). Fig. 1(c) visualizes the evolution of damage within the hard network (see Movie S1 in [47] for the complete evolution). In the initial elastic regime (A), the network remains pristine. Shortly before yielding (B), stochastic damage begins to emerge. At the yielding point (C), the coalescence of damages triggers crack initiation in the hard network. As the crack propagates across the sample cross-section, a necking band forms and the stress drops from $\sigma_y$ to $\sigma_p$ (D). The plateau regime (E) corresponds to axial propagation of the necking zone; here, the hard network is extensively fragmented (see [47]), and the fully elastic load is carried by the intact soft network. At the onset of strain hardening $\varepsilon_{harde}$ , the necking has developed through the entire sample. In the hardening regime (F), the hard network is fully fragmented, and the soft network carries the total load until final rupture (G).

Beyond reproducing this typical DN response, the model also reveals a brittle-to-ductile (BTD) transition (see Supplemental Material [47]): ductile DN behavior emerges only when the failure stress of the soft network ($\sigma_s^f$) exceeds that of the hard network ($\sigma_h^f$), whereas failure remains brittle when $\sigma_s^f < \sigma_h^f$. This BTD transition defines a mechanical phase boundary set by the failure-stress balance between the two networks, consistent with previous findings [30,31]. More strikingly, in the ductile case [Fig.1(b)], the yield strain distinctly exceeds the fracture strain of the standalone hard network, indicating that the soft network delays crack initiation in the hard network. The same trend is also observed in brittle DN systems (see Supplemental Material [47]).

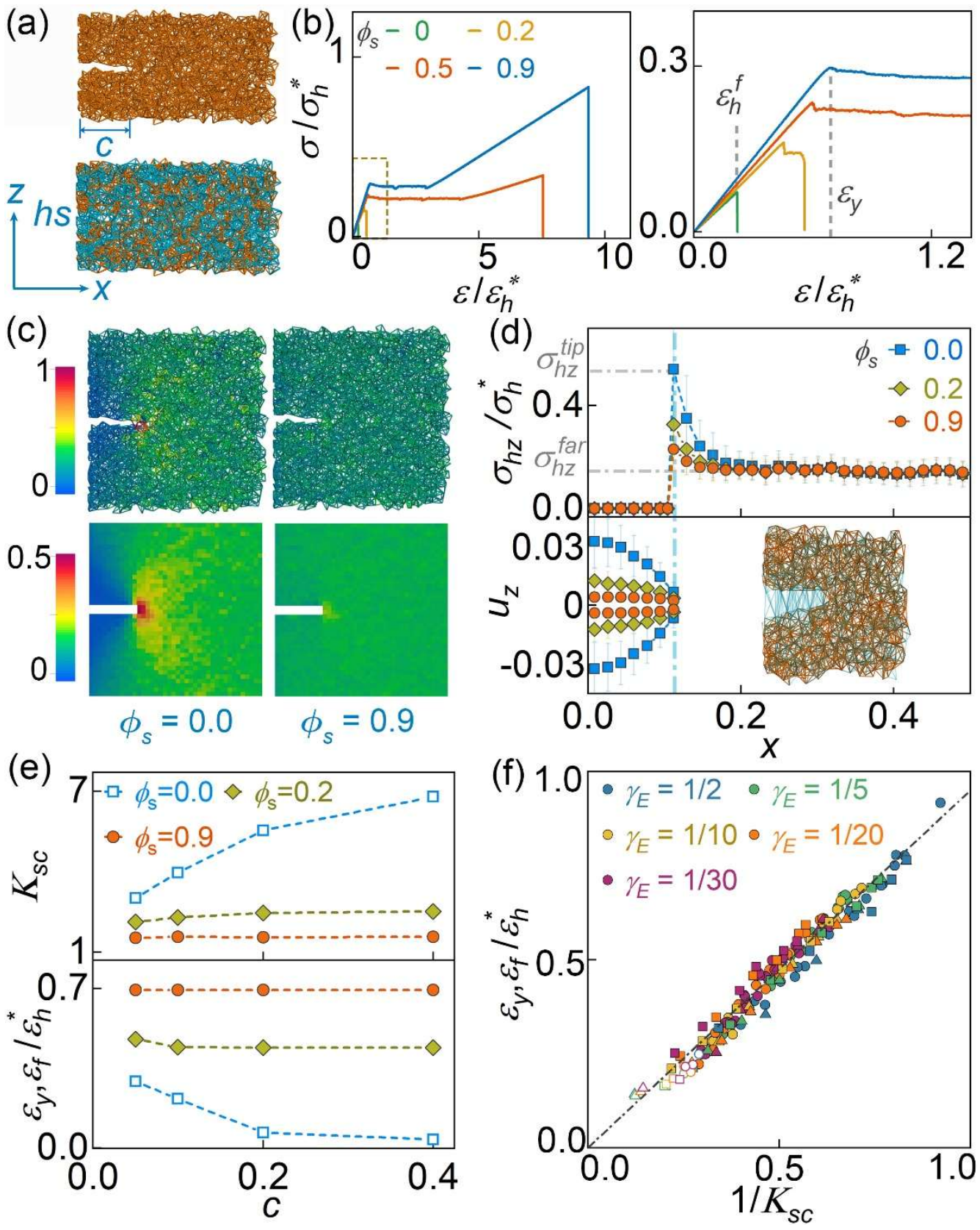

**Fig.2. Defect screening and the microscopic origin of yield enhancement. (a)** Schematic of a single-edge notched hard network (*h*) with notch length *c* = 0.1, and a double network (*hs*). Hard and soft networks are rendered in brown and blue, respectively. **(b)** Stress-strain curves for varying $\phi_s$ at fixed $\gamma_E = 1/10, \gamma_W = 24, \phi_h = 0.4$. Right: enlarged yield region. **(c)** Top: Hard-network tensile stress ($\sigma_{hz}/\sigma_h^*$) for $\phi_s = 0$ (left) and $\phi_s = 0.9$ (right) at $\varepsilon_h^f$. Bottom: Corresponding spatially averaged stress fields in the x-z plane [47]. **(d)** Top: Stress profiles within the hard network ahead of the notch tip, defining the stress concentration factor $K_{sc} = \sigma_{hz}^{tip}/\sigma_{hz}^{far}$ at $\varepsilon_h^f$ [see Supplemental Material [47] for details]. Bottom: Crack opening displacement (COD) profiles. **(e)** $K_{sc}$ (top) and yield strain $\varepsilon_y/\varepsilon_h^*$ (bottom) versus notch length *c* for various $\phi_s$. **(f)** Yielding and fracture strains $\varepsilon_y, \varepsilon_f/\varepsilon_h^*$ versus $1/K_{sc}$ for varying $\phi_s$ and $\gamma_E$. Circles, squares, and triangles denote $c = 0.1$, 0.2 and 0.4, respectively. Dashed line: linear fit through the origin with slope of 0.95. In (**e**, **f**), Fracture strains ($\varepsilon_f$) of brittle samples are shown by open symbols.

A fundamental question then arises: By what physical mechanism does the soft network shield the hard network and delay the onset of fracture? Motivated by recent experiments on pre-yield damage growth in DN gels [54], we introduce an edge notch of length *c* into the hard network at the mid-plane (*z* = 1), while keeping the soft network unnotched [Fig. 2(a)]. Under uniaxial tension, the system undergoes a BTD transition as $\phi_s$ increases at fixed $\phi_h = 0.4$ and $c = 0.1$ [Fig. 2(b)].

Fig. 2(c) displays thin x–z slices centered at the notch (top) and the corresponding spatially averaged stress fields (bottom) for $\phi_s = 0$ and $0.9$, captured at the fracture strain of the standalone hard network $\varepsilon_h^f$. Only the hard network is rendered, colored by its normalized tensile stress, $\sigma_{hz}/\sigma_h^*$. In the single hard network case ($\phi_s = 0$), the notch induces a pronounced stress concentration and intense strain localization. Adding a dense soft network significantly inhibits this localization, indicating effective defect screening, consistent with experimental mechanophore observations in multiple-network elastomers [33]. To quantify this shielding effect, Fig.2(d, top) plots the profile of $\sigma_{hz}/\sigma_h^*$ extracted from the stress fields (Fig. 2(c, bottom)) along the notch line. From this profile, we define a stress concentration factor $K_{sc} = \sigma_{hz}^{tip}/\sigma_{hz}^{far}$ as the ratio between the spatially averaged notch-tip stress ($\sigma_{hz}^{tip}$) to the far-field stress ($\sigma_{hz}^{far}$). From the perspective of linear elastic fracture mechanics [47], $K_{sc}$ is loading-level independent in the linear regime and serves as a dimensionless measure of local stress amplification. The associated crack-opening displacement (COD) profile $u_z$ of the hard network is plotted in Fig. 2(d, bottom). Both $K_{sc}$ and COD decrease systematically with increasing $\phi_s$ and $\gamma_E$ (see Supplemental Material [47]), demonstrating that soft-network bridging constrains notch opening and suppresses stress concentration.

We next examine how this screening depends on notch length. In the bare hard network, $K_{sc}$ is highly sensitive to *c*, as expected for a brittle material. Adding a soft network markedly lowers $K_{sc}$, and progressively suppresses this sensitivity as $\phi_s$ increases. For sufficiently dense soft networks ($\phi_s = 0.9$), $K_{sc}$ approaches unity and becomes independent of *c*, indicating the elimination of notch-sensitivity. The hard-network failure strain ($\varepsilon_f$) or the yield strain ($\varepsilon_y$) in ductile DN samples, follows the opposite trend [Fig.2(e, bottom)], suggesting an inverse relation with $K_{sc}$. Fig. 2(f) distills this behavior into a unified scaling law: the failure strain for both ductile and brittle regimes collapses onto a single master curve when plotted against the inverse stress-concentration factor, $1/K_{sc}$, across broad ranges of $\phi_s$, $\phi_h$, $\gamma_E$, and *c*. The resulting linear scaling, $\varepsilon_y,\ \varepsilon_f \approx \varepsilon_h^*/K_{sc}$, demonstrates that the local defect screening quantitatively governs the macroscopic strain at which the hard-network crack initiates.

As $K_{sc}$ approaches the full-screening limit ($K_{sc} \to 1$), the pre-existing notch no longer induces sufficient strain localization to selectively trigger crack initiation. Hard-network rupture then ceases to be notch-localized and instead occurs stochastically throughout the sample, fundamentally altering the spatial morphology of failure. To probe this transition, we construct two representative DN systems (Fig. 3(a)): yielding followed by stable necking (localized) and yielding with homogeneous deformation (delocalize). The corresponding deformation and damage patterns are visualized in Fig. 3(b).

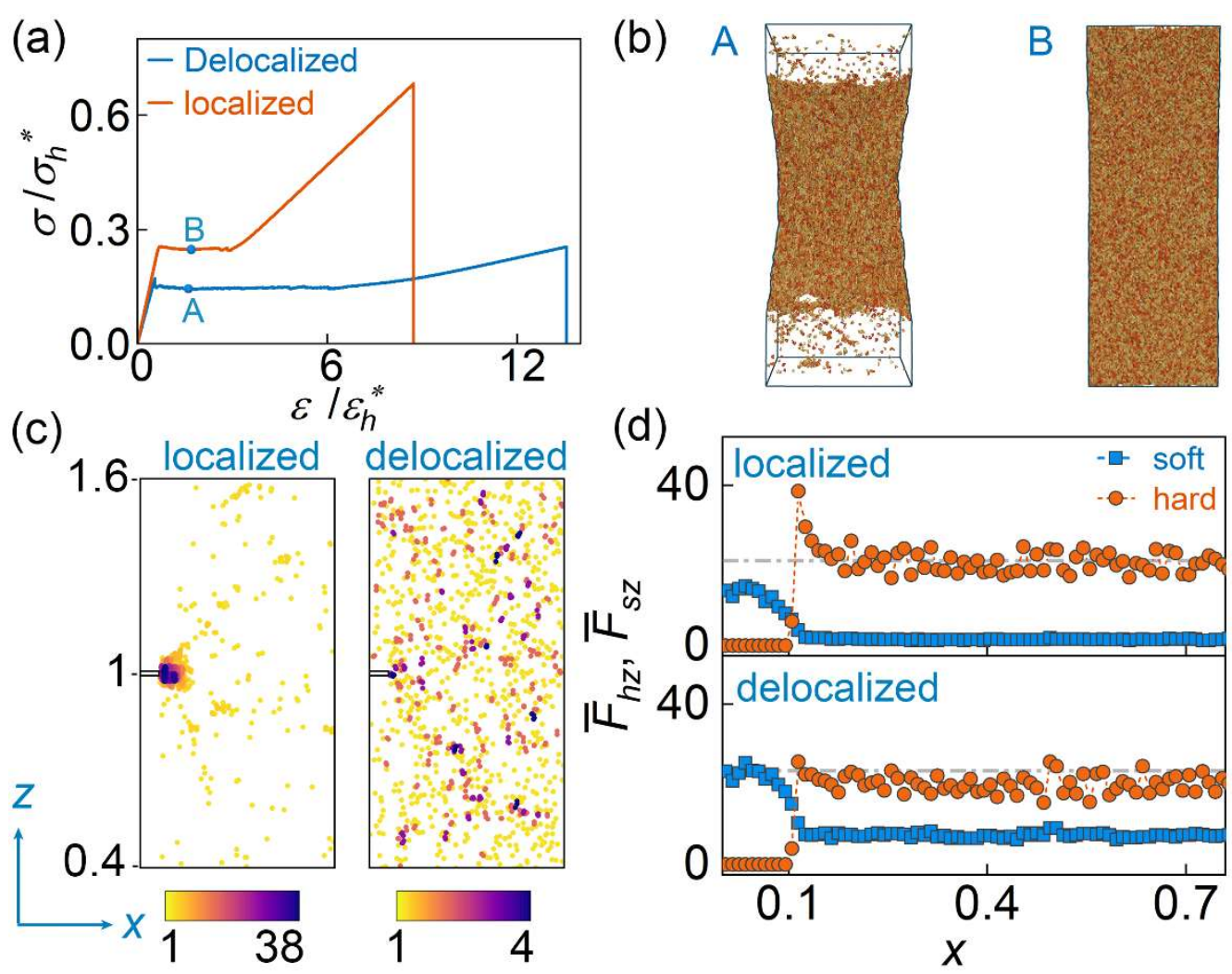

**Fig. 3. Transition from localized to delocalized hard-network damage in LHS-DN. (a)** Stress-strain response for two distinct deformation modes of DN systems: yielding with necking (localized damage) and yielding characterized by homogeneous deformation (delocalized damage). **(b)** Snapshots (A, B) illustrating the corresponding sample morphologies and hard-network damage pattern at the labels in (a). Brown markers indicate fractured hard elements. **(c)** Spatial density of broken hard elements at yielding, with color indicating the number of broken elements. **(d)** Load-bearing profiles ($\bar{F}_{hz}, \bar{F}_{sz}$) along the notch plane for a pre-notched ($c$ =0.1) hard network at a strain of $\varepsilon_h^f$ for the corresponding hard networks. Network parameters: $\phi_h = 0.35$, $\phi_s = 0.6$, $\gamma_w = 24$; $\gamma_E = 1/30$ (localized); $\gamma_E = 1/10$ (delocalized).

To elucidate the mechanics underlying these distinct modes, we analyze load redistribution around the defect using the notched geometry of Fig. 2. Fig. 3(c) shows the spatial density distribution of broken hard elements at yielding for the two DN systems and Fig. 3(d) shows the corresponding load-bearing profiles ($\bar{F}_{hz}, \bar{F}_{sz}$) of the constituent hard and soft networks along the notch plane, evaluated at a strain $\varepsilon_h^f$ of the corresponding hard networks [47]. These results identify inter-network load transfer near the defect as the key mechanical pathway by which soft-network bridging enables defect screening. In the localized case [Fig. 3(c), left], hard-network damage remains concentrated near the notch and evolves into a necking band. Correspondingly, the soft network carries much less load than the hard network along the notch line [Fig. 3(d), top]. Hard-network fracture thus creates a local load deficit that the soft network cannot sufficiently compensate, triggering a macroscopic stress drop and the subsequent necking. In the delocalized case [Fig. 3(c), right], broken hard elements are spatially dispersed, indicating that damage no longer localizes at the initial defect. Consistently, the soft network nearly fully compensates the load shedding caused by hard-network failure [Fig. 3(d), bottom], thereby suppressing localization and promoting spatially delocalized damage. These analyses reveal that the stress overshoot at the yield point and the subsequent necking are intrinsically correlated with the efficiency of inter-network load transfer.

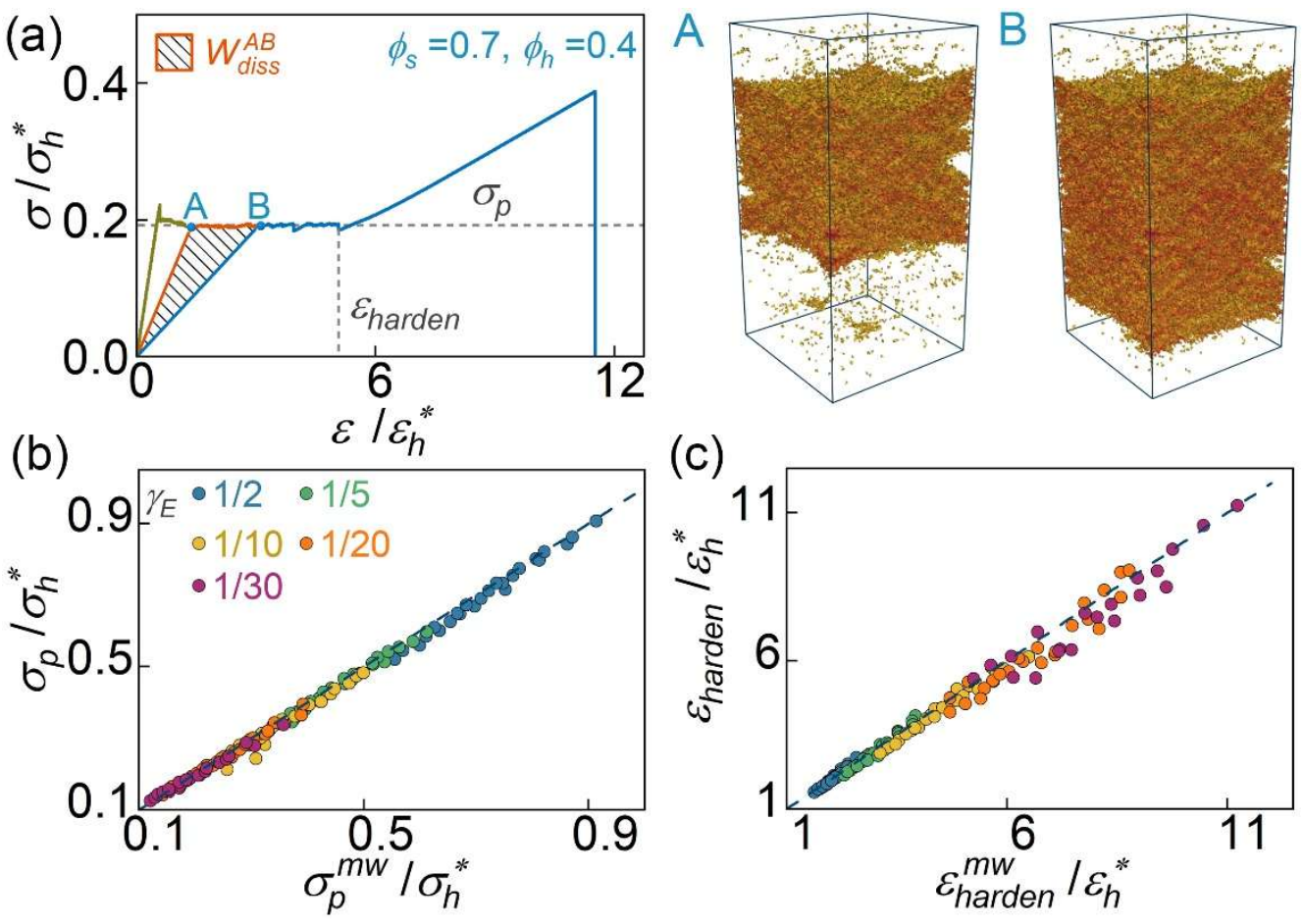

**Fig.4. Energetic origin of the necking plateau and hardening. (a)** Stress–strain response of a representative DN sample exhibiting stable necking ($\phi_s = 0.7, \phi_h = 0.4, \gamma_E = 1/20, \gamma_w = 24$). The shaded area $W_{diss}$ denotes the energy dissipated during steady-state neck propagation between A and B. The hardening strain $\varepsilon_{harden}$, indicates the completion of the phase transition. The right panel shows snapshots at A and B for deformation and corresponding hard-network damage. Brown markers indicate ruptured hard elements. **(b)** Plateau stress $\sigma_p$ and **(c)** hardening strain $\varepsilon_{harden}$ measured in simulations plotted against the corresponding Maxwell predictions $\sigma_p^{mw}$ and $\varepsilon_{harden}^{mw}$, respectively. Data are shown for varying $\phi_s$ and $\gamma_E$ (following the legend in b). The dashed lines in (b, c) denote the identity line ($y = x$).

Finally, we identify the physical principle that selects the plateau stress, $\sigma_p$ in the necking regime. We address this by treating the necking process as a stress-driven phase transition [39]. As illustrated in Fig. 4(a), snapshots A and B visualize the steady-state propagation of the neck, during which hard-network elements rupture sequentially at a localized front. This process converts the material from an unnecked phase, with effective modulus $E_{DN} \approx E_h(\phi_h + \gamma_E\phi_s)$ to a necked phase in which the hard network is fully damaged and no longer contributes to the stiffness. The necked phase is therefore carried entirely by the intact soft network, with modulus $E_{neck} \approx \phi_s E_s$ [47].

To determine the driving stress $\sigma_p$, we employ a generalized Maxwell construction that accounts for irreversible dissipation [55]. Necking is thus described as the coexistence of an unnecked state at strain $\varepsilon_{unneck}$ and a necked state at strain $\varepsilon_{neck}$, both coexisting at the common stress $\sigma_p$. The propagation condition requires the drop in potential-energy density ($\Pi(\varepsilon)$) to balance the energy dissipated per unit volume, $W_{diss}$, i.e., $\Pi(\varepsilon_{unneck}) - \Pi(\varepsilon_{neck}) = W_{diss}$. Here, $W_{diss} = \chi\phi_h W_h^*$ accounts for both hard-network rupture and the associated elastic relaxation [30,51,53]. We evaluate the dimensionless factor $\chi$ as the ratio of the accumulated dissipation (the shaded area $W_{diss}^{AB}$ in Fig. 4(a)) to the total intrinsic energy of the newly broken hard elements. In the linear-elastic framework, the coexistence condition dictates $\sigma_p = E_{unneck}\,\varepsilon_y = E_{neck}\varepsilon_{neck}$, which yields the Maxwell prediction for the plateau stress [47]

$$\sigma_p^{mw} = \sigma_h^* \sqrt{\chi \gamma_E \phi_s (\phi_h + \gamma_E \phi_s)} \tag{1}$$

Fig. 4(b) shows excellent agreement between $\sigma_p^{mw}$ from Eq. 1 and the simulated $\sigma_p$ measured in across different $\gamma_E$ and varying $(\phi_h, \phi_s)$. The same picture also predicts the end of necking: once the hard network is extensively fragmented, the macroscopic response is governed by the necked phase, giving the hardening-strain estimate $\varepsilon_{harden}^{mw} = \sigma_p^{mw}/E_{neck}$. As shown in Fig.4(c), this estimate closely matches $\varepsilon_{harden}$ extracted from simulations. These results identify the necking plateau as the Maxwell stress of a dissipation-regulated phase transition, selected by the balance between potential-energy release and rupture dissipation. Although derived for the minimal LHS-DN model, this criterion points to a general selection principle for hard-soft double networks.

In summary, we developed a minimal LHS–DN model that captures the signature mechanical response of tough DN gels. We identify defect screening by soft-network bridging as the microscopic origin of toughening. The inverse stress-concentration factor, $1/K_{sc}$, quantifies this screening and scales linearly with macroscopic failure strain, directly linking local defect shielding to global strengthening. We further elucidated that the transition from localized damage (necking) to delocalized damage of the hard network is mediated by the efficiency of inter-network load transfer, whereas the brittle-to-ductile transition is dictated by the failure-stress balance between the constituent networks. In the localized regime, the macroscopic necking plateau is identified as the Maxwell stress of a dissipation-regulated phase transition, selected by the energetic balance between potential-energy release and rupture dissipation. Such rich emergent behavior from a minimal model reveals an intrinsic principle: the hard–soft contrast underpins defect screening and the suppression of stress localization. Ultimately, the load-transfer capacity and failure-stress competition synergistically determine the macroscopic failure mode. These findings provide a predictive framework for designing ultra-tough architectures and motivate future studies on spatial correlation lengths and material nonlinearity in complex networks.

**Acknowledgments**

The authors acknowledge the support from JSPS International Research Fellow in Japan (Grant No. 23KF0002). J.P.G. thanks the JSPS KAKENHI (No. 22H04968, 22K21342).

*Data availability*—The data that support the findings of this article are not publicly available. The data are available from the authors upon reasonable request.

## Supplemental Material for

# Defect screening and load transfer in minimal hard–soft double networks

Fucheng Tian[1]*, Feixue Lu[2], Katsuhiko Sato[3], Liangbin Li[4], Bin Li[5], Jian Ping Gong [1,2]*

* Fucheng Tian

Email: tianfc@sci.hokudai.ac.jp

* Jian Ping Gong

Email: gong@sci.hokudai.ac.jp

**This PDF file includes:**

Supporting text

Figures S1 to S9

Legends for Movies S1 to S4

SI References

**Other supporting materials for this manuscript include the following:**

**Movies S1 to S4**

## S1. Model

**S1.1. LHS-DN model.** The core idea of the proposed LHS–DN model is presented in the main text. Here we provide the implementation details. We consider a bar-shaped specimen ($l_0 \times l_0 \times 2l_0$), representing the gauge region of a uniaxial tensile sample, which is discretized into $N_{tot}$ tetrahedral elements using a three-dimensional Delaunay triangulation with characteristic mesh size $l_e = l_0/60$. This mesh resolution is chosen to ensure that the microscopically heterogeneous architecture exhibits a macroscopically homogeneous response (statistical self-averaging), as discussed in the next section. To generate the hard and soft networks as random element sets on this common background mesh, we introduce two logical arrays $I_h(e)$ and $I_s(e)$ of length $N_{tot}$. Each element $e$ is thus characterized by two binary membership indicators $I_h(e)$, $I_s(e) \in \{0,1\}$, specifying whether it belongs to the hard and soft networks, respectively. The hard network is constructed by randomly selecting $N_h$ distinct elements from $\{1,\ldots,N_{tot}\}$ and setting $I_h(e) = 1$ for these elements and $I_h(e) = 0$ otherwise. The resulting hard-network relative density is $\phi_h = N_h/N_{tot}$. Independently, the soft network is generated by randomly choosing $N_s$ elements, assigning $I_s(e) = 1$ to the selected elements and $I_s(e) = 0$ for all others, with soft-network relative density $\phi_s = N_s/N_{tot}$. In this discrete representation, elements with $(I_h, I_s)$=(1,0) carry only the hard-network response, elements with (0,1) carry only the soft-network response, elements with (1,1) contain both networks and bear load in parallel, and elements with (0,0) are voids that contribute neither stiffness nor energy, acting as topological defects. Different random seeds for the sampling of $I_h$ and $I_s$ generate microscopically distinct yet macroscopically equivalent network realizations for fixed $\phi_h$ and $\phi_s$.

Under quasi-static loading, the displacement field $\boldsymbol{u}$ of the LHS-DN system satisfies

$$\nabla \cdot (\boldsymbol{\sigma}_h + \boldsymbol{\sigma}_s) = \mathbf{0} \tag{S1}$$

with body forces neglected. For each phase $\alpha \in \{h, s\}$, we adopt a linear-elastic constitutive law. The small-strain tensor is defined as

$$\boldsymbol{\varepsilon} = \frac{1}{2}(\nabla\boldsymbol{u} + \nabla\boldsymbol{u}^T) \tag{S2}$$

and the corresponding strain-energy density reads

$$W_\alpha(\boldsymbol{\varepsilon}) = \frac{1}{2}\boldsymbol{\varepsilon}: \mathbb{C}_{\boldsymbol{\alpha}}: \boldsymbol{\varepsilon} \tag{S3}$$

where $\mathbb{C}_{\boldsymbol{\alpha}}$ represents the fourth-order elastic tensor defined by the elastic modulus $E_\alpha$ and Poisson's ratio $\nu$ (taken identical for both phase). To capture the brittle fracture of individual network strands, we introduce an irreversible, element-wise damage variable $d_\alpha \in \{0,1\}$ for each phase. The Cauchy stress of phase $\alpha$ in element $e$ is then given by

$$\boldsymbol{\sigma}_\alpha(e) = [1 - d_\alpha(e)]\frac{\partial W_\alpha}{\partial \boldsymbol{\varepsilon}} = [1 - d_\alpha(e)]\mathbb{C}_{\boldsymbol{\alpha}}: \boldsymbol{\varepsilon} \tag{S4}$$

Network rupture evolves according to a local energy-based criterion. For each element, we compute the strain-energy density of phase $\alpha$ at its single Gauss integration point and track the historical maximum of this quantity. Once the historical maximum strain-energy density at that Gauss point exceeds the prescribed threshold $W_\alpha^*$, the corresponding stiffness/energy contribution in this element is irreversibly removed

$$d_\alpha(e) = \begin{cases} 1, & W_\alpha(\boldsymbol{\varepsilon}(e,\tau)) \geq W_\alpha^* \\ 0, & \text{otherwise.} \end{cases} \tag{S5}$$

The history dependence in Eq. S5 ensures that $d_\alpha$ is monotonic, i.e., network failure is irreversible and a broken element never recovers load-bearing capacity. Based on the above construction, the effective elastic tensor of element $e$ can be derived as

$$\mathbb{C}_e(I_h, I_s, d_h, d_s) = I_h(e)[1 - d_h(e)]\mathbb{C}_h + I_s(e)[1 - d_s(e)]\mathbb{C}_s \tag{S6}$$

The displacement field $\boldsymbol{u}$ is defined on the global node set. Nodes that are connected to both hard and soft elements act as inter-network junctions. Load transfer between networks is therefore mediated intrinsically by displacement compatibility and nodal force balance at these shared nodes, without introducing any additional interface constitutive law.

The numerical implementation employs a displacement-controlled quasi-static loading protocol. The displacement increment $\nabla u_z$ at the top and bottom surfaces is adaptively adjusted within the range $10^{-4}\sim 10^{-7} l_0$, depending on the internal fracture activity. At each load step, the equilibrium state is determined by an inner equilibrium–failure fixed-point iteration. First, Eq. (S1) is solved using the current damage configuration $\{d_h, d_s\}$ to obtain a trial elastic field. Elements that newly satisfy the energetic fracture criterion [Eq. (S6)] are then identified and their damage variables are updated ($d_\alpha \to 1$). The global stiffness matrix is subsequently reassembled, and the system is re-equilibrated at the same prescribed loading state. This update loop is repeated until the set of damaged elements becomes stationary, i.e., no further failures occur at that load level. In this way, the scheme naturally captures cascading fracture events driven by stress redistribution, while maintaining a robust quasi-static loading conditions.

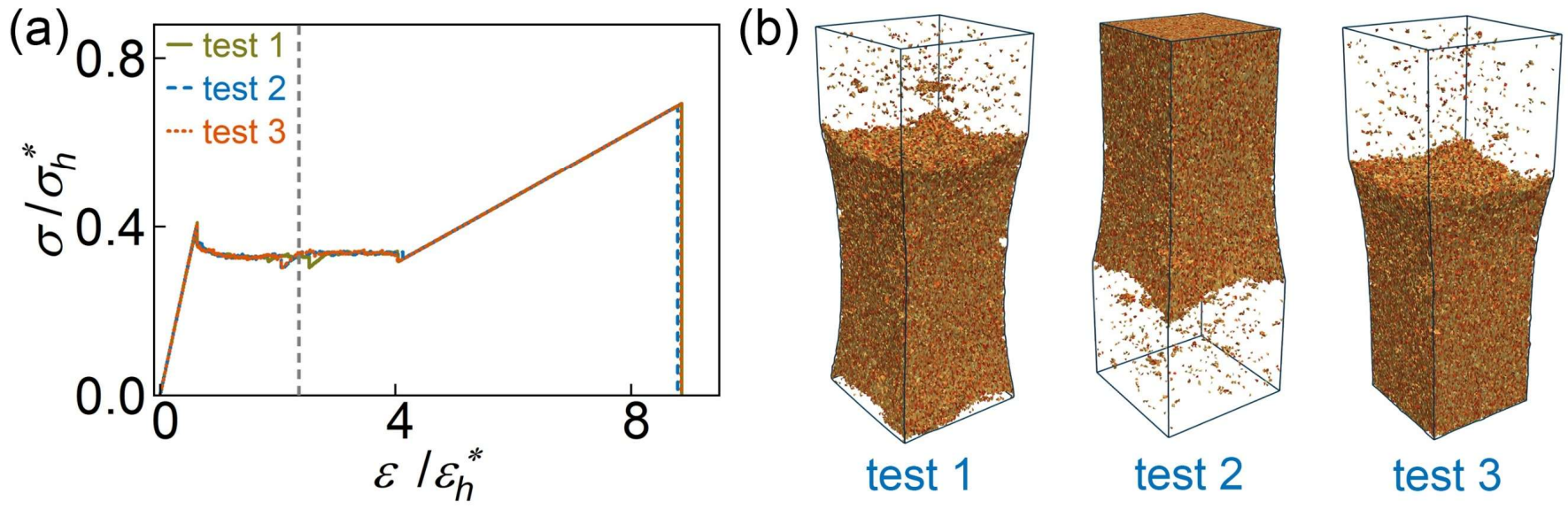

**Fig. S1. Macroscopic homogeneity of the LHS–DN model.** (a) Nominal stress–strain curves for three independent LHS–DN realizations with $\phi_h = 0.6$, $\phi_s = 0.8$, $\gamma_E = 1/10$ and $\gamma_W = 24$, generated using different random seeds. (b) Corresponding hard-network damage patterns at the same applied strain for the three realizations.

**S1.2. Macroscopic Homogeneity**. The LHS–DN model inherits a key feature of its SH–*com* prototype: macroscopic homogeneity emerges when the element size satisfies $l_e/l_0 > 50$. As stated above, we adopt $l_e = l_0/60$ to ensure this condition. To validate this, we perform three independent realizations of the representative DN system shown in Fig. 1 of the main text (with $\phi_h = 0.6$, $\phi_s = 0.8$, $\gamma_E = 1/10$ and $\gamma_W = 24$), using different random seeds. The resulting nominal stress–strain curves in Fig. S1(a) almost coincide, indicating that the macroscopic response is essentially insensitive to the microscopic randomness in the network realization. Conversely, the hard-network damage patterns at the same applied strain [Fig. S1(b)] exhibit distinct spatial morphologies, reflecting the inherent structural randomness. This confirms that the LHS-DN system is microscopically heterogeneous yet statistically homogeneous at the specimen scale, supporting the robustness of our findings.

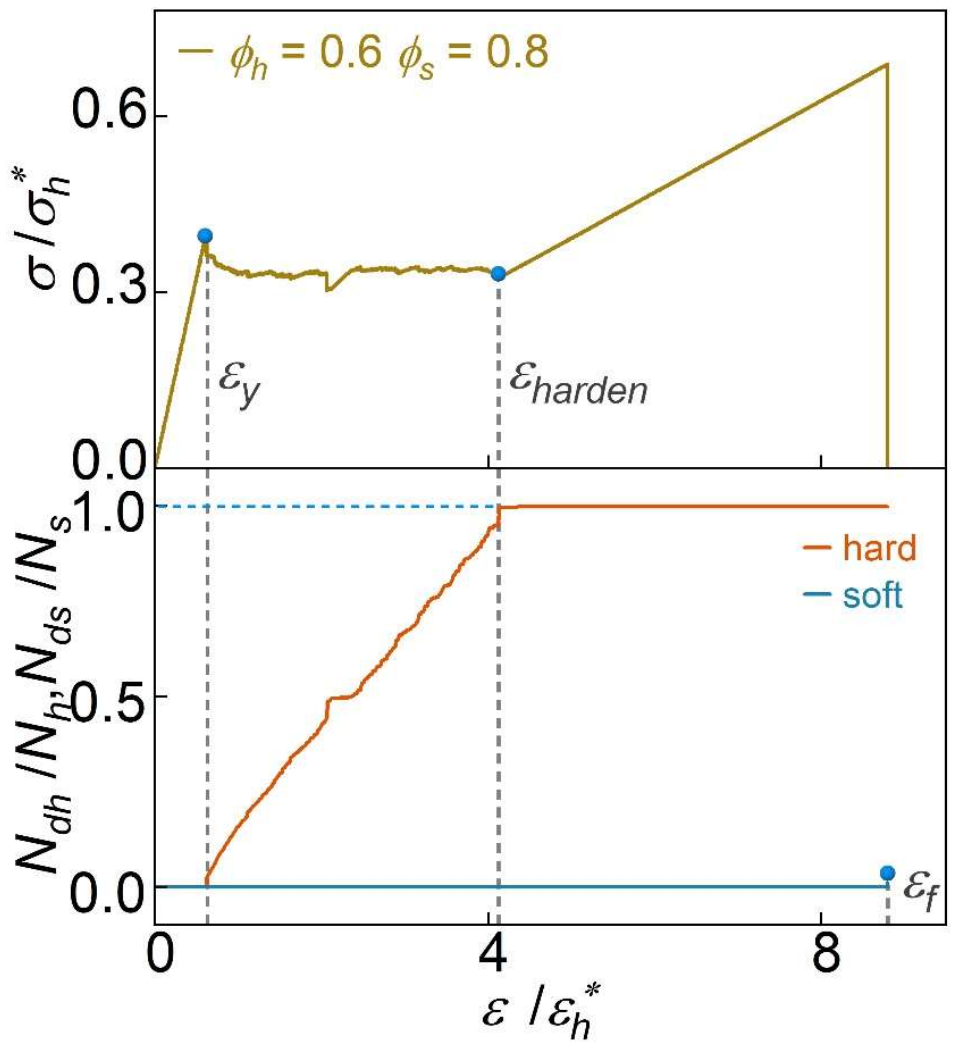

**Fig. S2. Internal fracture evolution in a representative DN system.** Top: representative stress–strain curve of the DN system [same case as Fig. 1(b) in the main text]. Bottom: evolution of the fractions of broken hard and soft elements, $N_{dh}/N_h$ and $N_{ds}/N_s$, during loading. Before yielding, both networks remain nearly intact. Hard-network fracture initiates at yielding and accumulates rapidly during necking, reaching near-complete failure by the end of the necking regime. The subsequent hardening response is then carried primarily by the soft network. Soft-network fracture begins only near the terminal stage, and complete rupture is triggered by the breaking of only a very small fraction of soft elements.

### S1.3 Sequential internal fracture in a representative DN system

To complement the internal damage evolution of the DN system shown in Fig. 1 of the main text, we track the fractions of broken hard ($N_{dh}/N_h$) and soft elements $N_{ds}/N_s$ during loading, as shown in Fig. S2. Here, $N_{dh}$ and $N_{ds}$ denote the numbers of broken hard and soft elements, respectively, while $N_h$ and $N_s$ are the corresponding total numbers of hard and soft elements. The upper panel shows the corresponding stress–strain response, while the lower panel plots $N_{dh}/N_h$ and $N_{ds}/N_s$ as functions of strain.

In initial linear elastic regime, both networks remain intact. Upon yielding, fracture initiates in the hard network, and the fraction of broken hard elements $N_{dh}/N_h$ rises rapidly throughout the necking regime. By the end of the necking plateau, the sacrificial hard network is nearly fully fragmented ($N_{dh}/N_h \to 1$). Consequently, the subsequent strain-hardening regime is sustained entirely by the intact soft network. In contrast, damage in the soft network remains negligible over most of the deformation history and emerges only as the DN system approaches ultimate failure. Complete rupture is ultimately triggered by the breaking of only a very small fraction of soft elements. These results further support the internal damage evolution of DN systems described in the main text.

**S2. Influence of specimen geometry**

In the main text, the bar-shaped specimen was chosen with an initial height $h_0 = 2l_0$. To assess the influence of specimen geometry, we vary the height while keeping all material parameters fixed. Specifically, we consider $h_0 = l_0$, $2l_0$ and $3l_0$. Fig. S3(a) shows the resulting normalized stress–strain curves ($\sigma/\sigma_h^* - \varepsilon/\varepsilon_h^*$). The three curves almost coincide over the entire loading history, indicating that the normalized macroscopic response is essentially insensitive to the specimen height within this range. Fig. S3(b) further visualizes the microscopic damage evolution by selecting a representative loading state in the necking regime (indicated by the vertical dashed line in Fig. S3(a)). For clarity, only the broken hard-network elements at this strain are displayed and highlighted in orange. While the variation in specimen height alters the global aspect ratio, the macroscopic mechanical response remains essentially unchanged, showing that our main conclusions do not depend on the specific specimen geometry.

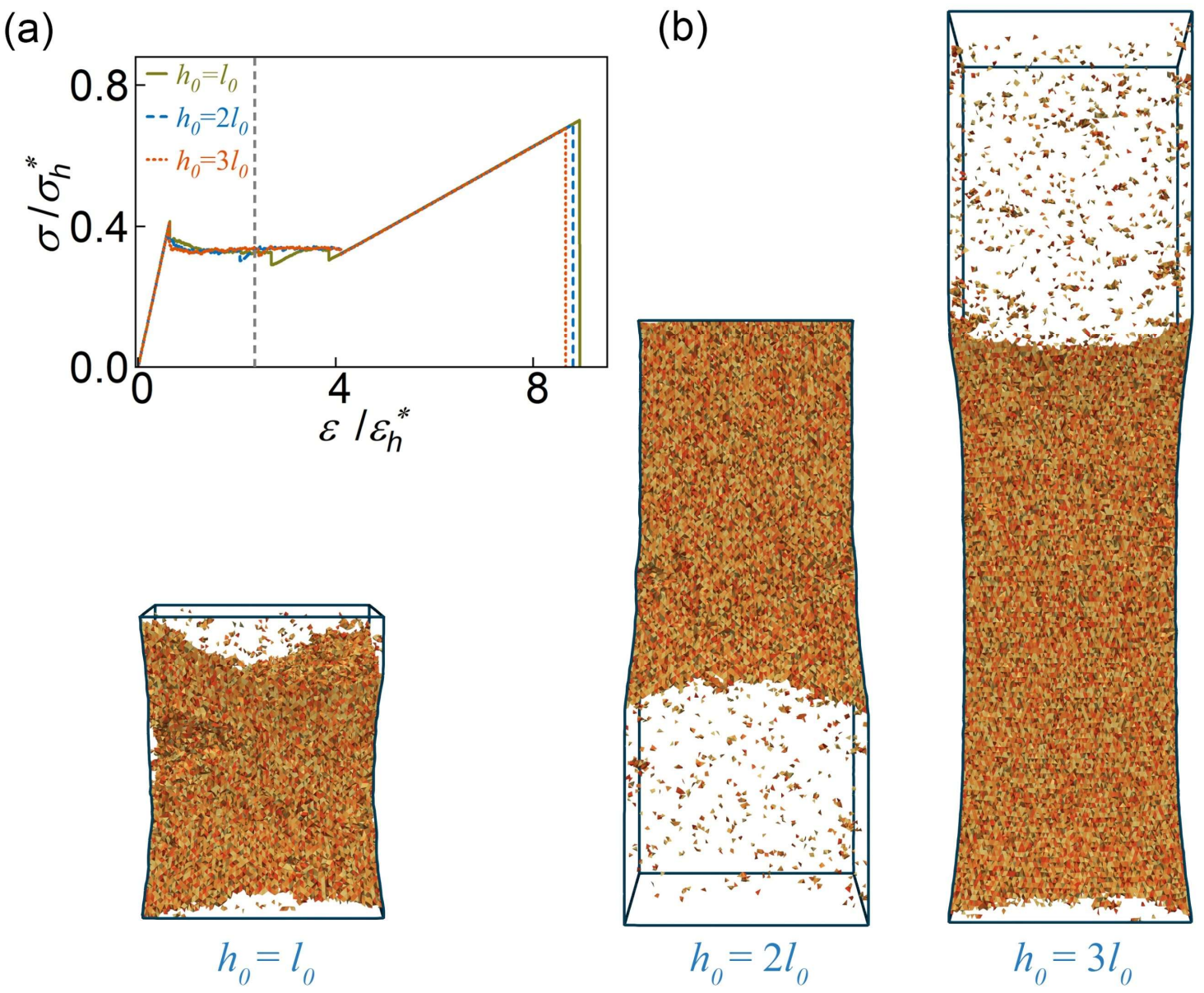

**Fig. S3. Influence of specimen height on macroscopic response and damage morphology.** (a) Normalized stress–strain curves ($\sigma/\sigma_h^* - \varepsilon/\varepsilon_h^*$) for specimens with initial heights $h_0 = l_0$, $2l_0$ and $3l_0$. The three curves nearly overlap, showing that the normalized mechanical response is insensitive to the specimen geometry. The vertical dashed line indicates the loading state used for the snapshots in panel (b). (b) Deformed configurations at the loading state marked in (a) for $h_0 = l_0$, $2l_0$ and $3l_0$. The broken hard networks at this state are rendered in orange.

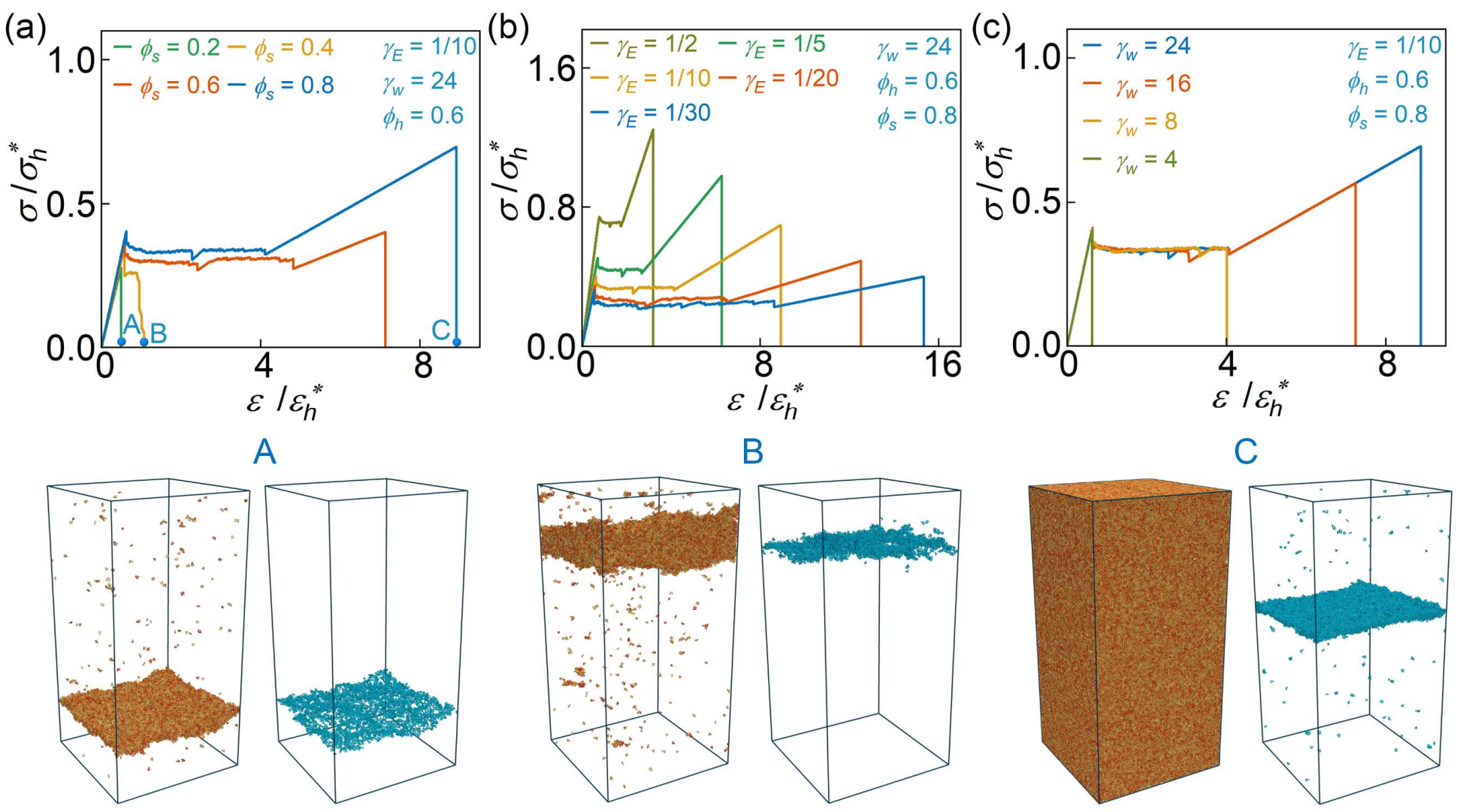

**Fig. S4. Effect of network parameters on the mechanical behavior of LHS-DN systems.** (a) Stress–strain curves for different soft-network densities $\phi_s$ with fixed $\phi_h = 0.6$, $\gamma_E = 1/10$ and $\gamma_W = 24$. The mechanical response evolves from brittle ($\phi_s = 0.2$) to ductile ($\phi_s = 0.6, 0.8$) behavior, with $\phi_s = 0.4$ representing a transition case. (b) Effect of modulus contrast $\gamma_E$ (fixed $\phi_h = 0.6$, $\phi_s = 0.8$ and $\gamma_W = 24$). Smaller $\gamma_E$ leads to larger failure strain but lower yield and plateau stresses. (c) Effect of critical energy-density contrast $\gamma_W$ (fixed $\phi_h = 0.6$, $\phi_s = 0.8$ and $\gamma_E = 1/10$). $\gamma_W$ primarily regulates the final fracture strain, while leaving the yielding and plateau response essentially unchanged. Bottom panels (A–C): fracture morphologies corresponding to points A–C marked in (a). The broken hard and soft networks are rendered as orange and blue, respectively. (A) Brittle system with highly localized hard-network damage. (B) Transition case with a moderately thickened damage zone. (C) Toughened DN system with widely distributed hard-network rupture.

## S3. Macroscopic mechanical response and the brittle-to-ductile transition

### S3.1. Effect of network parameters on the mechanical behavior

In this section, we show how varying network parameters modifies the macroscopic response of the LHS–DN system. While the analysis in Fig. 2 of the main text was performed at a hard-network density of $\phi_h = 0.4$, here we adopt $\phi_h = 0.6$. The results exhibit qualitatively identical trends to those in the main text, confirming the robustness of the underlying mechanisms.

In Fig. S4(a), we first examine the effect of the soft-network volume fraction $\phi_s$ at fixed modulus and critical strain energy density contrasts ($\gamma_E = 1/10, \gamma_W = 24$). At a low soft-network density ($\phi_s = 0.2$), the response remains brittle. As $\phi_s$ is increased to 0.6 and 0.8, the mechanical response of the DN system evolves from brittle to ductile, with the fracture strain becoming nearly an order of magnitude larger than that at $\phi_s = 0.2$. This enhanced toughness arises from soft-network bridging across defects in the fragmented hard network, which mitigates stress concentrations and delays localization. The intermediate case $\phi_s = 0.4$, where the failure stresses of the hard and soft networks are comparable, represents a transition state with mild toughening and a slightly thickened damage zone. The brittle-to-ductile (BTD) transition observed here is governed by the balance of failure stresses between the two networks, as elucidated in the subsequent Fig. S5.

Fig. S4(b) displays the influence of the modulus contrast $\gamma_E$ at fixed $\phi_h = 0.6$, $\phi_s = 0.8$ and $\gamma_W = 24$. Decreasing $\gamma_E$ (making the soft network more compliant) leads to a systematic increase in the fracture stretch, while both the yield and plateau stresses decrease. This behavior is consistent with experimental trends reported for tough DN gels [1]. Within the LHS–DN framework, this follows because macroscopic failure in the toughened regime is governed by the soft network: reducing its modulus lowers the macroscopic stress scale but increases the critical strain required to trigger intrinsic soft-network failure. Next, we examine the role of $\gamma_W$. Varying $\gamma_W$ primarily affects the ultimate fracture strain, while leaving the pre-fracture response essentially unchanged (see Fig. S4(c)). In particular, both the yielding behavior and the necking plateau are insensitive to $\gamma_W$, which motivates fixing $\gamma_W$ when analyzing the microscopic origin of yielding in Fig. 2 of the main text.

The lower panels (A–C) in Fig. S4 visualize the damage morphologies at failure for the three representative cases in Fig. S4(a) (orange: broken hard elements; blue: broken soft elements). In the brittle case (A), the damage zone in the hard network remains strongly localized, acting as the crack path in conjunction with the rupturing soft network. Conversely, the ductile case (C) exhibits pervasive fragmentation of the hard network throughout the bulk, indicating massive energy dissipation via distributed rupture. The transition case (B) shows a mildly thickened damage zone, consistent with its intermediate toughening effect.

### S3.2. Brittle-to-ductile transition

Section S3.1 has demonstrated how individual network parameters drive the BTD transition. Here, we establish a general physical criterion governing this transition. In both brittle and ductile LHS-DN systems, we selectively deactivated either the hard or soft network to isolate their individual mechanical contributions [2], as shown in Fig.S5(a). The resulting isolated responses reveal that brittle behavior arises when the hard network's failure stress exceeds that of the soft network, i.e., $\sigma_h^f > \sigma_s^f$. When this inequality is reversed, $\sigma_h^f < \sigma_s^f$, the DN system enters a ductile regime. The corresponding final failure patterns in Fig. S5(b) further highlight this distinction: brittle systems exhibit highly localized hard-network damage zone, whereas ductile systems display spatially extended hard-network failure. This stress criterion is generalized in Fig. S5(c) by scanning a broad parameter space ($\phi_h$, $\phi_s$, $\gamma_E$, $\gamma_W$). In this plot, the normalized DN failure strain ($\varepsilon_f/\varepsilon_s^f$) is plotted against the failure stress ratio $\sigma_h^f/\sigma_s^f$, with the failure-strain ratio $\varepsilon_s^f/\varepsilon_h^f$ encoded by color. A sharp transition clearly emerges near $\sigma_h^f/\sigma_s^f \approx 1$. The absence of any systematic trend in the color map corroborates that the BTD transition is governed by the failure-stress balance between the networks [2,3], rather than by their failure-strain contrast.

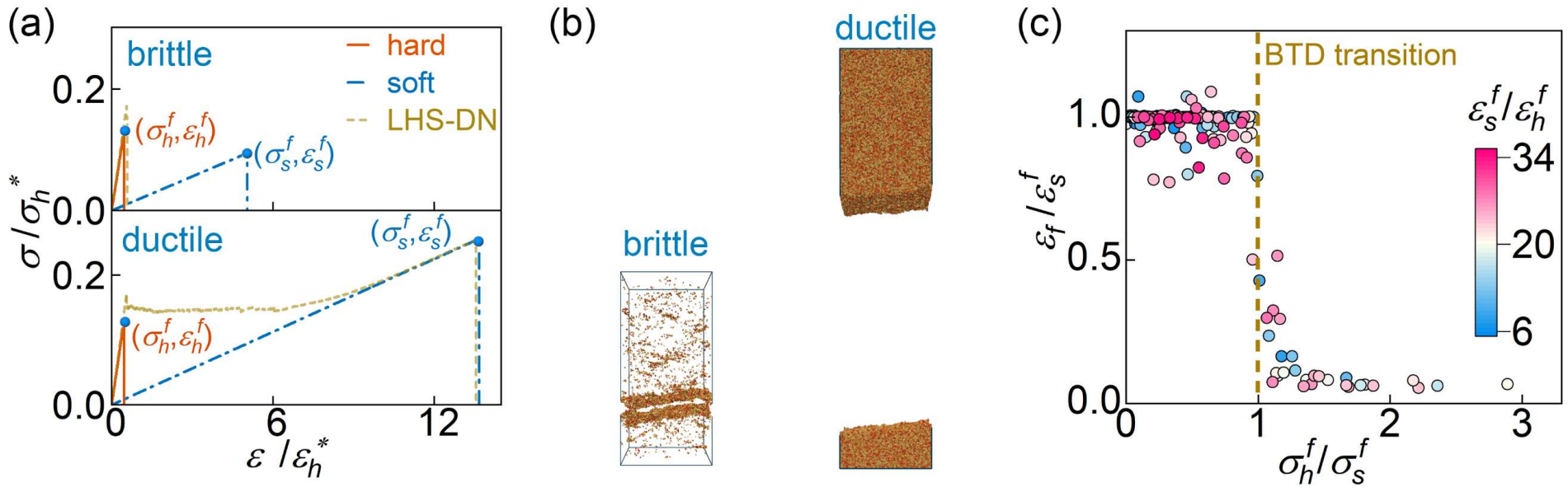

**Fig. S5. Criteria for brittle to ductile (BTD) transition in LHS-DN system.** (a) Isolated stress–strain curves of the hard and soft networks, obtained by selectively deactivating the soft and hard phases. Upper: in brittle DN ($\phi_s = 0.35, \gamma_E = 1/30, \gamma_w = 4$); lower: in ductile DN ($\phi_s = 0.35, \gamma_E = 1/30, \gamma_w = 24$). (b) Final failure patterns for the brittle (left) and ductile (right) cases. Damaged hard-network elements are highlighted in brown, revealing highly localized hard-network failure in the brittle case, versus spatially extended hard-network damage in the ductile case. (c) DN failure strain ratio $\varepsilon_f/\varepsilon_s^f$ versus stress ratio $\sigma_h^f/\sigma_s^f$ shows a sharp BTD boundary near $\sigma_h^f/\sigma_s^f \approx 1$. The color indicates the failure-strain ratio of single networks $\varepsilon_s^f/\varepsilon_h^f$.

## S4. Stress concentration factor, continuum mechanics interpretation, and numerical evaluation

### S4.1 Continuum mechanics interpretation of $K_{sc}$

In Fig. 2(f) of the main text, we demonstrate a robust linear scaling between the macroscopic yield or fracture strain $\varepsilon_y$ and the inverse stress–concentration factor $1/K_{sc}$ (i.e., $\varepsilon_y \propto 1/K_{sc}$). Here, $K_{sc}$ is not the classical mode-I stress intensity factor (SIF), but a finite-scale, dimensionless measure of crack-tip stress amplification.

For a sharp mode-I crack in linear elastic fracture mechanics (LEFM) [4], the near-tip tensile stress ahead of the crack satisfies

$$\sigma_{zz}(r, 0) \approx \frac{K_I}{\sqrt{2\pi r}} \tag{S7}$$

where $r$ is the distance from the crack tip and $K_I$ is the mode-I stress intensity factor. For a finite-width specimen containing a single-edge crack of length $c$,

$$K_I = Y(c/L_0)\, \sigma_{far}\sqrt{\pi c} \tag{S8}$$

where $\sigma_{far}$ is the remote tensile stress, $L_0$ is the specimen width, and $Y(c/L_0)$ is the geometry factor. Thus, in the continuum sharp-crack limit, the crack-tip amplification is governed by both the geometry ratio $Y(c/L_0)$ and the crack length $c$.

In the present minimal model, however, the characteristic element size $l_e$ also serves as the intrinsic structural scale of the network representation. We therefore do not evaluate the stress concentration at the mathematical singularity ($r \to 0$), but at the first resolved structural scale near the notch tip. This is analogous in spirit to the Theory of Critical Distances (TCD) [5,6], in which failure is assessed from the stress evaluated over a finite characteristic length rather than from the singular hot-spot value.

Evaluating the near-tip stress at this structural scale ( $r \sim l_e$ ) yields

$$\sigma_{tip} \approx \frac{Y\left(\frac{c}{L_0}\right)\sigma_{far}\sqrt{\pi c}}{\sqrt{2\pi l_e}} = Y(c/L_0)\,\sigma_{far}\sqrt{c/2l_e} \tag{S9}$$

Accordingly, the corresponding dimensionless stress-concentration factor scales as

$$K_{sc} \equiv \frac{\sigma_{tip}}{\sigma_{far}} \approx Y(c/L_0)\sqrt{c/2l_e} \tag{S10}$$

Eq. (S10) shows explicitly that, for a fixed notch geometry and a given structural length scale $l_e$, $K_{sc}$ is independent of the remote loading magnitude in the linear regime.

### S4.2 Numerical evaluation of $K_{sc}$ in the discrete LHS–DN system

The proposed LHS-DN model is built from linear tetrahedral elements, with hard and soft phases randomly assigned in space. Consequently, the local stress in the hard network exhibits strong element-wise fluctuations, rendering raw cell data unsuitable for defining a reproducible stress profile ahead of the notch. To obtain a well-defined smooth stress field, we project the hard-network normal stress in the loading direction, $\sigma_{hz}$, onto the *x-z* plane and perform a statistical average along the y direction. The *x-z* plane is partitioned into square bins of edge length $l_{bin}$. For each bin, we identify the set of $N_h^y$ hard elements whose centroids lie inside the square footprint along the *y*-axis. normalized bin-averaged stress is then computed as

$$\sigma_{hz}/\sigma_h^* = \bar{F}_{hz}/N_h^y, \quad \bar{F}_{hz} = \sum_{i=1}^{N_h^y} \frac{\sigma_{hz}^{g(i)}}{\sigma_h^*} \tag{S11}$$

where $\sigma_{hz}^{g(i)}$ represents the hard-network normal stress at the Gauss point of the *i*-th hard element in the loading direction. The summation term represents the integrated load-bearing capacity of the hard network within the bin, denoted here as $\bar{F}_{hz}$. This quantity corresponds directly to the 'load-bearing' metric analyzed in Fig. 3 of the main text. An identical averaging procedure is applied to the soft network to compute its corresponding load-bearing intensity $\bar{F}_{sz}$ and normalized average stress $\sigma_{sz}/\sigma_h^*$. Sampling these bin-averaged stress values along the notch plane (along *x* at the notch plane) yields a smooth one-dimensional (1D) stress profile ahead of the notch (see Fig. 2(d) in the main text). The stress-concentration factor is defined as $K_{sc} = \sigma_{hz}^{tip}/\sigma_{hz}^{far}$, where $\sigma_{hz}^{tip}$ and $\sigma_{hz}^{far}$ denote the peak tip stress and far-field stress extracted from this coarse-grained profile.

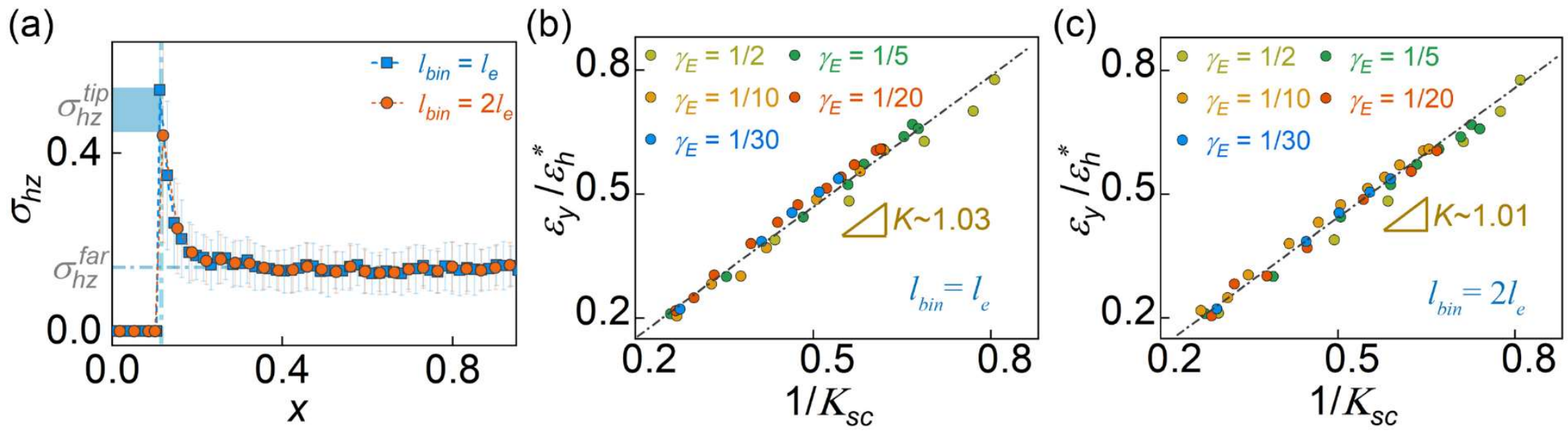

**Fig. S6. Effect of square-bin size on the stress-concentration factor $K_{sc}$ and the yield strain scaling.** (a) Normalized hard-network normal stress in the loading direction, $\sigma_{hz}/\sigma_h^*$ , along the notch front for bin sizes $l_{bin} = l_e$ and $l_{bin} = 2l_e$. Increasing $l_{bin}$ slightly smooths the tip stress peak due to spatial averaging over a larger spatial region, while the far-field plateau stress remains essentially unchanged. (b), (c) Normalized yield strain $\varepsilon_y/\varepsilon_h^*$ plotted against the inverse stress-concentration factor $1/K_{sc}$ for $l_{bin} = l_e$ (b) and $l_{bin} = 2l_e$ (c). In both cases the data show a clear linear scaling $\varepsilon_y \propto 1/K_{sc}$.

As discussed in Section 4.1, the characteristic element size $l_e$ serves as a natural length scale for binning. An excessively large bin size would artificially smooth out the peak stress gradient, whereas an overly small bin would capture insufficient elements for averaging, rendering the result sensitive to microscopic fluctuations. We therefore set $l_{bin} = l_e$ for all calculations presented in the main text. To verify that the scaling $\varepsilon_y \propto 1/K_{sc}$ not an artefact of this particular choice, we repeated the analysis using a coarser bin size, $l_{bin} = 2l_e$. Fig. S6(a) compares the resulting stress profiles. As expected, the larger bin size reduces the apparent peak stress $\sigma_{hz}^{tip}$ due to spatial averaging, while the far-field stress level $\sigma_{hz}^{far}$ remains unchanged. This quantitative reduction in peak stress, however, does not alter the qualitative physics. Figs. S6(b) and S6(c) plot the normalized yield strain $\varepsilon_y/\varepsilon_h^*$ as a function of $1/K_{sc}$ for $l_{bin} = l_e$ and $l_{bin} = 2l_e$, respectively, across a range of system parameters. In both cases, the data exhibit a clear linear dependence, although the specific slope varies slightly. This confirms that the connection between defect screening and macroscopic yielding identified in the main text is intrinsic rather than a numerical artefact.

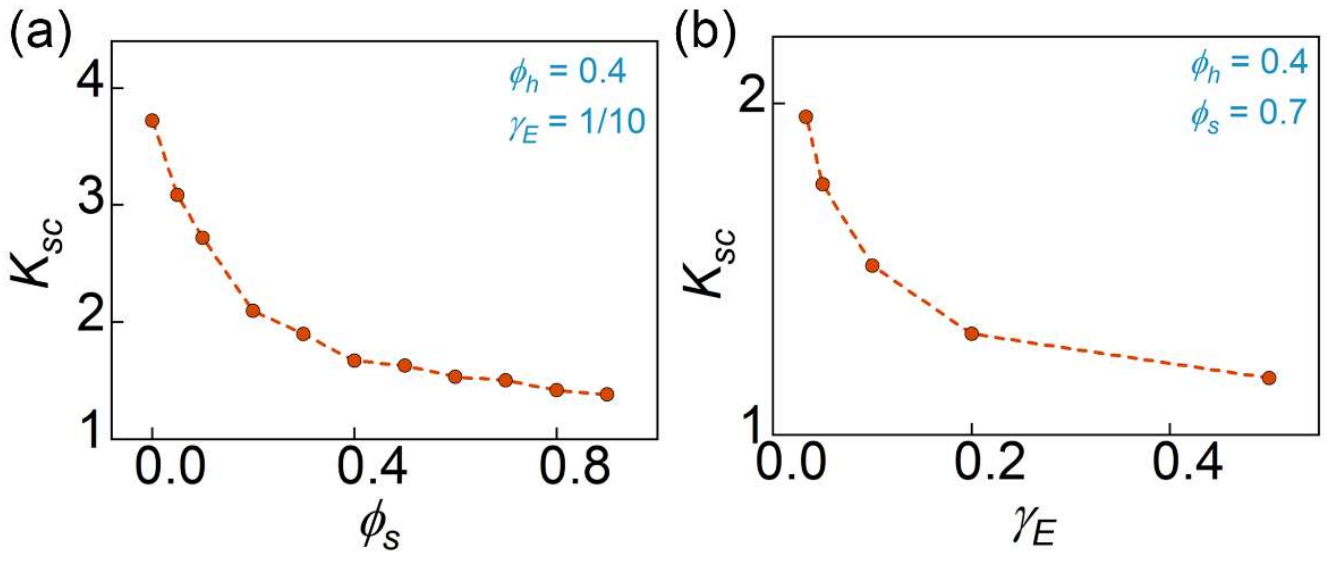

**Fig. S7. Influence of soft-network architecture on the stress concentration factor $K_{sc}$.** (a) $K_{sc}$ as a function of the soft-network density $\phi_s$ for $\phi_h = 0.4$ and $\gamma_E = 1/10$. (b) $K_{sc}$ versus modulus

contrast $\gamma_E$ at fixed $\phi_h = 0.4$ and $\phi_s = 0.7$. In these tests, we fix $\gamma_w = 24$, as it has negligible impact on the pre-fracture response (e.g., yielding).

### S4.3. Dependence of $K_{sc}$ on soft-network architecture

To provide additional insight into the master scaling in Fig. 2(f), we explore how $K_{sc}$ varies with the soft-network density and modulus. By definition, a larger $K_{sc}$ corresponds to stronger stress amplification at the defect and hence weaker defect screening. Fig. S7(a) shows that increasing the soft-network density $\phi_s$ (fixed $\phi_h = 0.4$, $\gamma_E = 1/10$) leads to a monotonic reduction in $K_{sc}$ that gradually levels off as $\phi_s$ increases. This trend reflects that a denser soft network bridges defect more effectively, thereby mitigating stress concentrations. Fig. S7(b), obtained at fixed $\phi_h = 0.4$ and $\phi_s = 0.7$, reveals that increasing the modulus contrast $\gamma_E$ follows a trend similar to increasing $\phi_s$, likewise leading to a reduction in $K_{sc}$. In other words, a stiffer soft network screens defect more effectively by sharing the load carried by the hard skeleton. However, stronger defect screening does not imply enhanced macroscopic toughening. As shown in Fig. S4(b), a more compliant soft network (lower $\gamma_E$), despite exhibiting higher stress concentrations and earlier yielding, sustains significantly larger strains before ultimate failure.

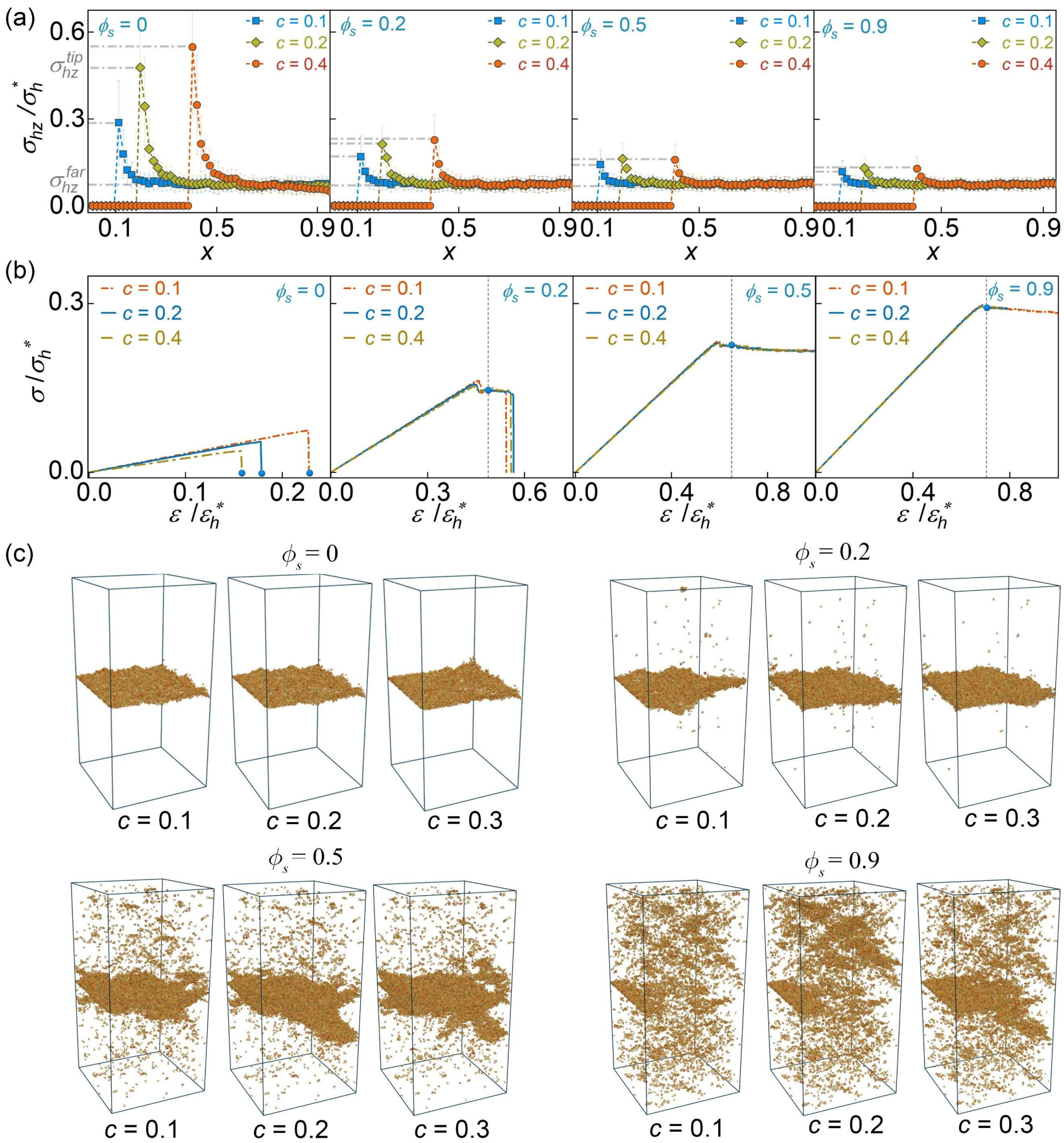

**Fig. S8. Evolution of the local stress profile, macroscopic response, and damage morphology with hard-network notch length for different soft-network densities.** (a) Normalized hard-network tensile-stress profile ahead of the notch, $\sigma_{hz}/\sigma_h^*$, for three notch lengths, $c = 0.1$, 0.2, and 0.4, shown for $\phi_s = 0$, 0.2, 0.5, and 0.9. (b) Corresponding normalized macroscopic stress-strain curves. (c) Internal damage patterns at the marked points in (b), where broken hard-network elements are highlighted in brown. With increasing $\phi_s$, both the local stress concentration and the macroscopic response become progressively less sensitive to notch length, consistent with enhanced defect screening by the soft network. Here, $\phi_h = 0.4$, $\gamma_w = 24$, and $\gamma_E = 1/10$ are fixed.

### S4.4. Evolution of $K_{sc}$ and $\varepsilon_y$ with hard-network notch length

In the main text, the dependence of the stress-concentration factor $K_{sc}$ and the fracture or yielding strain $\varepsilon_y$ on the hard-network notch length $c$ is summarized in Figs. 4(e) and 4(f). Here,

Fig. S8 provides the underlying full-field results for different soft-network densities, $\phi_s = 0, 0.2$, 0.5 and 0.9. For each $\phi_s$, Fig. S8(a) shows the normalized hard-network tensile-stress profile ahead of the notch, $\sigma_{hz}/\sigma_h^*$, for three notch lengths, $c = 0.1$, 0.2, and 0.4, extracted at a reference strain corresponding to the macroscopic failure strain of the $c = 0.4$ case. Fig. S8(b) shows the corresponding normalized macroscopic stress-strain curves. The internal damage patterns at the marked points in Fig. S8(b) are displayed in Fig. S8(c), where broken hard-network elements are highlighted in orange.

For the bare hard network, $\phi_s = 0$, increasing $c$ strongly amplifies the notch-tip stress $\sigma_{hz}^{tip}$, whereas the far-field stress $\sigma_{hz}^{far}$ remains nearly unchanged. As a result, $K_{sc}$ increases markedly with notch length, while the macroscopic failure strain decreases substantially, indicating pronounced notch sensitivity. The corresponding damage patterns show brittle fracture initiated from the pre-existing notch. Upon introducing a soft network, this notch sensitivity is progressively suppressed as $\phi_s$ increases. For $\phi_s = 0.2$ and 0.5, the variation in both the local stress profile and the macroscopic response with $c$ is already significantly reduced, and the damage patterns in both cases show the formation of a necking band. The key difference is that the $\phi_s = 0.2$ system remains brittle, whereas the $\phi_s = 0.5$ system becomes ductile, consistent with the different failure-stress contrast between the two networks discussed above in connection with the brittle-to-ductile transition. For the dense soft network, $\phi_s = 0.9$, the stress profiles for different notch lengths exhibit nearly identical peak values ($\sigma_{hz}^{tip}$) and only very weak stress concentration, indicating that $K_{sc} \to 1$. The corresponding stress-strain curves are nearly coincident, and the damage patterns become spatially delocalized. The corresponding stress-strain curves are nearly coincident, and the damage patterns become spatially delocalized, indicating that the initial hard-network notch no longer triggers strain localization. These supplementary results further support the main-text conclusion that sufficiently efficient load transfer to the soft network suppresses notch sensitivity by screening the hard-network defect.

**S5. Maxwell construction incorporating irreversible dissipation**.

In this section, we provide a detailed derivation of the generalized Maxwell construction used in the main text. From an energetic perspective, this theory provides semi-analytical predictions for the plateau stress ($\sigma_p^{mw}$) and the associated hardening strain ($\varepsilon_{harden}^{mw}$) in the LHS-DN model.

**S5.1. Energetic Formulation.** During the necking plateau, the DN specimen can be viewed as the coexistence of two (approximately) homogeneous states under a common tension stress $\sigma_p$: an unnecked phase characterized by effective modulus $E_{unneck}$ and strain $\varepsilon_{unneck}$, and a necked phase characterized by an effective modulus $E_{neck}$ and strain $\varepsilon_{neck}$. These two regions are separated by a transformation front. Under quasi-static conditions, front propagation is governed by a generalized Maxwell construction that accounts for irreversible dissipation. Let us consider a material element with unit volume undergoing a transformation from the unnecked state to the necked state while the applied tensile stress is held at $\sigma_p^{mw}$. The incremental energy balance per unit volume reads

$$\sigma_p^{mw} \mathrm{d}\varepsilon = dW(\varepsilon) + dW_{diss} \tag{S12}$$

where $W(\varepsilon)$ is the recoverable strain-energy density and $W_{diss}$ is the accumulated irreversible dissipation due to hard-network rupture and associated elastic relaxation. Integrating Eq. S12 from $\varepsilon_{unneck}$ to $\varepsilon_{neck}$ gives

$$\sigma_p^{mw}(\varepsilon_{neck} - \varepsilon_{unneck}) - (W(\varepsilon_{neck}) - W(\varepsilon_{unneck})) = W_{diss} \tag{S13}$$

where $W_{diss}$ is the dissipated energy density for the **unnecked** → **necked** transformation. Rearranging Eq. (S13) leads to the propagation condition in terms of the potential energy density, defined as $\Pi(\varepsilon) = W(\varepsilon) - \sigma_p^{mw}\varepsilon$, where the term $\sigma_p^{mw}\varepsilon$ represents the mechanical work per unit volume supplied by the external load. The energy balance (S13) thus becomes

$$\Pi(\varepsilon_{unneck}) - \Pi(\varepsilon_{neck}) = W_{diss} \tag{S14}$$

which is the generalized Maxwell condition quoted in the main text.

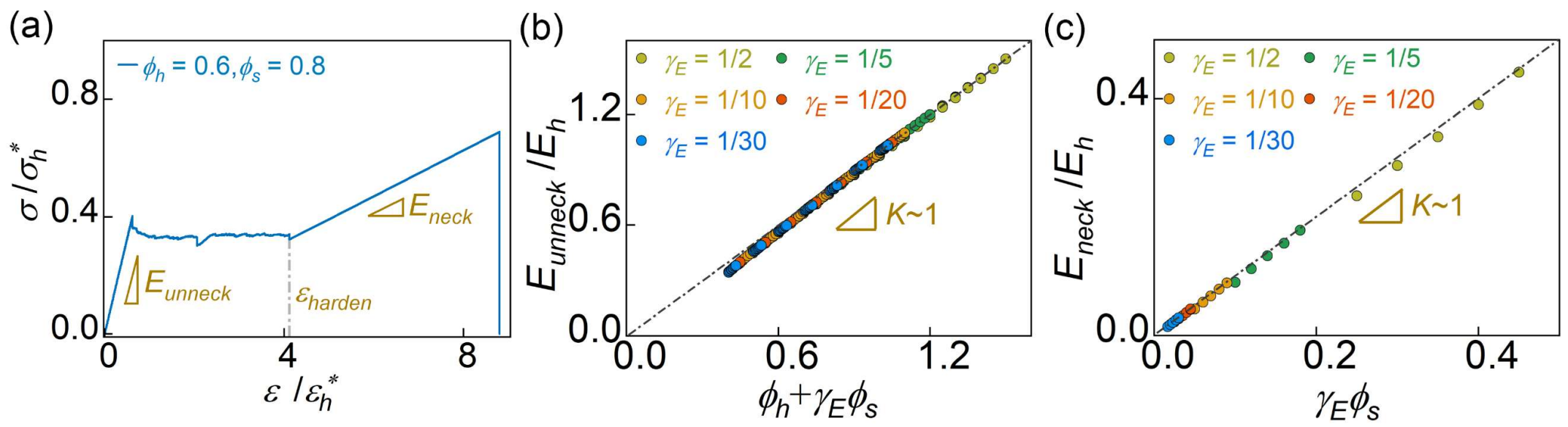

**Fig. S9. Extraction and scaling of effective moduli.** (a) Representative macroscopic stress–strain curve illustrating the definitions of the pristine modulus $E_{unneck}$ and the necked modulus $E_{neck}$. (b) Normalized pristine modulus $E_{unneck}/E_h$ plotted against the mixture parameter $\phi_h + \gamma_E\phi_s$. The data collapse onto the line $y = x$ (dash line), showing the linear rule of mixtures. (c) Normalized pristine modulus $E_{neck}/E_h$ plotted against the soft-network contribution $\gamma_E\phi_s$. The linear scaling indicates that the necked state is dominated by the soft phase.

**S5.2. Specialization to the LHS–DN model.** In the LHS–DN model, both coexisting states are treated as linearly elastic. Under the plateau stress $\sigma_p^{mw}$, the axial strains of the unnecked state and the necked state are therefore

$$\varepsilon_{unneck} = \frac{\sigma_p^{mw}}{E_{unneck}}, \qquad \varepsilon_{neck} = \frac{\sigma_p^{mw}}{E_{neck}} \tag{S15}$$

Substituting these relations into Eq. (S9) yields

$$\frac{\sigma_p^{mw\,2}}{2}\left(\frac{1}{E_{neck}} - \frac{1}{E_{unneck}}\right) = W_{diss} \tag{S16}$$

The effective moduli $E_{unneck}$ and $E_{neck}$ can be extracted directly from the simulated stress–strain curves by fitting the initial elastic regime and the post-plateau hardening regime, respectively, as illustrated in Fig. S9(a). Simulation results obtained by systematically varying ($\phi_s$ , $\phi_h$ , $\gamma_E$), shown in Figs. S9(b)-(c), suggest that

$$E_{unneck} \approx E_h(\phi_h + \gamma_E\phi_s), \qquad E_{neck} \approx \phi_s E_s \tag{S17}$$

These relations indicate that, in the LHS–DN system, the effective modulus is well described by a nearly linear superposition of the hard and soft contributions in the unnecked state, whereas the necked-state modulus is governed by the soft network. Consequently, the onset of hardening marks the point where the hard network is extensively fragmented, leaving the soft network to sustain the macroscopic load. It should be noted, however, that experimental observations often reveal a more complex scenario where the hard network forms spatially correlated clusters. Unlike the random distribution assumed here, these clusters can survive the necking phase and contribute to the strain-hardening response. Extending the LHS–DN model to account for such spatial correlations is the next step in the future.

**S4.3. Quantification of Energy Dissipation.** In the energy balance equation (Eq. S16), the term $W_{diss}$ represents the specific dissipated energy per unit volume during the transformation from the unnecked phase to the necked phase. Considering that the intrinsic energy density of the hard network per unit volume is given by $\phi_h W_h^*$, and accounting for the non-local elastic relaxation induced by hard-network rupture, we formulate the total specific dissipation energy density as

$$W_{diss} = \chi \phi_h W_h^* \tag{S18}$$

where $\chi$ is a dimensionless dissipation coefficient. Physically, $W_{diss}$ comprises both the intrinsic rupture energy of hard elements and the kinetic energy dissipated via elastic snap-back of the surrounding medium [2,7,8]. Therefore, the coefficient $\chi$ serves as an amplification factor (typically $\chi > 1$) that quantifies the ratio of this total energetic cost to the intrinsic bond failure energy.

We determine $\chi$ numerically by correlating the macroscopic energy dissipation with the microscopic network-breaking events observed in the simulation. Specifically, we select two distinct states (see Fig. 4 in the main text), A and B, located on the necking plateau (sufficiently separated to minimize statistical fluctuations). First, we calculate the macroscopic energy dissipated in the system $D_{macro}^{AB}$ between these two states. This corresponds to the specific dissipated work density $W_{diss}^{AB}$(the shaded area in Fig. 4(a) in the main text) multiplied by the total system volume $V_{tol}$, i.e., $D_{macro}^{AB} = W_{diss}^{AB} V_{tol}$. Meanwhile, we monitor the microscopic damage process and count the number of rupture hard networks between A and B, denoted as $N_{dh}^{AB}$. The corresponding intrinsic rupture-energy cost is given by $D_{int}^{AB} = N_{dh}^{AB} W_h^* V_e$, where $V_e$ is the volume of a single tetrahedral element. The dimensionless dissipation factor $\chi$ is then evaluated as the ratio of the actual macroscopic dissipation to the theoretical intrinsic energy cost

$$\chi = \frac{D_{macro}^{AB}}{D_{int}^{AB}} = \frac{W_{diss}^{AB} V_{tol}}{N_{dh}^{AB} W_h^* V_e} = \frac{N_{tot} W_{diss}^{AB}}{N_{dh}^{AB} W_h^*} \tag{S19}$$

with the geometric relation $V_{tol}/V_e \approx N_{tot}$. Eq. S19 thus provides a direct, statistically robust measurement of $\chi$ from the simulation data. The resulting semi-analytical predictions (Eq. 1 in the main text) exhibit excellent agreement with the simulation results, as demonstrated in Fig. 4 of the main text.

**Supplementary Movie Captions**

Movie S1 (separate file). Mechanical response and internal fracture evolution of a single hard network ($\phi_h = 0.6$) undergoing brittle failure. Broken hard elements are highlighted in orange.

Movie S2 (separate file). Mechanical response and internal fracture evolution of a toughened DN system exhibiting stable necking ($\phi_h = 0.6$, $\phi_s = 0.8$, $\gamma_E = 1/10, \gamma_w = 24$). Broken hard and soft elements are highlighted in orange and blue, respectively.

Movie S3 (separate file). Mechanical response and internal damage evolution of a toughened DN system exhibiting delocalized (homogeneous) internal fracture ($\phi_h = 0.3$, $\phi_s = 0.8$, $\gamma_E = 1/10, \gamma_w = 24$). Broken hard and soft elements are highlighted in orange and blue, respectively.

Movie S4 (separate file). Brittle-to-ductile (BTD) transition in the LHS–DN system. Brittle case: $\phi_s = 0.35, \gamma_E = 1/30, \gamma_w = 4$; ductile case: $\phi_s = 0.35, \gamma_E = 1/30, \gamma_w = 24$.